\documentclass[journal]{vgtc}                

\ifpdf
  \pdfoutput=1\relax                   
  \usepackage{graphicx}                
  \DeclareGraphicsExtensions{.pdf,.png,.jpg,.jpeg} 
\else
  \usepackage{graphicx}                
  \DeclareGraphicsExtensions{.eps}     
\fi%

\graphicspath{{figures/}{pictures/}{images/}{./}} 

\usepackage{microtype}                 
\PassOptionsToPackage{warn}{textcomp}  
\usepackage{textcomp}                  
\usepackage{mathptmx}                  
\usepackage{times}                     
\usepackage{cite}                      
\usepackage{tabu}                      
\usepackage{booktabs}                  

\usepackage{amsmath}
\usepackage{amsfonts}
\usepackage{amssymb}

\usepackage[usenames,dvipsnames,table]{xcolor}

\usepackage{microtype}

\usepackage{gincltex}
\usepackage{booktabs}
\usepackage{pifont}
\usepackage{wasysym}
\usepackage[utf8]{inputenc}
\usepackage{tablefootnote}

\usepackage[lined,boxed,commentsnumbered]{algorithm2e}

\usepackage{soul}

\usepackage[table]{xcolor}
\definecolor{c1}{rgb}{0,0,1}
\definecolor{c2}{rgb}{0,0.5,1}

\usepackage{tabularx}

\usepackage{gincltex}
\usepackage{booktabs}
\usepackage{pifont}
\usepackage{wasysym}
\usepackage[utf8]{inputenc}
\usepackage{tablefootnote}

\usepackage{tikz}

\newcommand{\pie}[1]{%
	\begin{tikzpicture}
	\draw (0,0) circle (0.7ex);\fill (0.7ex, 0) arc (0:#1:0.7ex) -- (0,0) -- cycle;
	\end{tikzpicture}%
} 

\onlineid{0}

\vgtccategory{Research}
\vgtcpapertype{please specify}

\title{Topology-Preserving Off-screen Visualization:\\Effects of Projection Strategy and Intrusion Adaption}

\author{Dominik J\"ackle, Johannes Fuchs, Harald Reiterer}
\authorfooter{
 \item All authors are with the University of Konstanz, Germany. E-mail: \{dominik.jaeckle, johannes.fuchs, harald.reiterer\}@uni-konstanz.de
}

\shortauthortitle{Biv \MakeLowercase{\textit{et al.}}: Global Illumination for Fun and Profit}

\abstract{
With the increasing amount of data being visualized in large information spaces, methods providing data-driven context have become indispensable.
\textit{Off-screen visualization} techniques, therefore, have been extensively researched for their ability to overcome the inherent trade-off between overview and detail. 
The general idea is to project off-screen located objects back to the available screen real estate.
Detached visual cues, such as halos or arrows, encode information on position and distance, but fall short showing the topology of off-screen objects.
For that reason, state of the art techniques integrate visual cues into a dedicated border region.
As yet, the dimensions of the navigated space are not reflected properly, which is why we propose to adapt the intrusion of the border pursuant to the position in space. 
Moreover, off-screen objects are projected to the border region using one out of two projection methods: \textit{Radial} or \textit{Orthographic}.
We describe a controlled experiment to investigate the effect of the adaptive border intrusion to the topology as well as the users' intuition regarding the projection strategy. 
The results of our experiment suggest to use the orthographic projection strategy for unconnected point data in an adaptive border design. 
We further discuss the results including the given informal feedback of participants.%
} 

\keywords{Overview Preservation, Off-screen Visualization, Projection Strategy, Adaptive Intrusion.}

\CCScatlist{ 
 \CCScat{K.6.1}{Management of Computing and Information Systems}%
{Project and People Management}{Life Cycle};
 \CCScat{K.7.m}{The Computing Profession}{Miscellaneous}{Ethics}
}

\renewcommand{\manuscriptnotetxt}{\vspace{1em}}

\vgtcinsertpkg

\begin{document}


\firstsection{Introduction}

\maketitle



\looseness=-1
The ever-growing collection of data demands efficient methods to navigate the information space. 
Geo-spatial observations, large scatterplots, and results of multivariate projections are prominent examples. 
A key task is to localize objects of interest and to put them into relation with each other. 
The limited screen real estate brings in the problem of tackling overview when exploring such large information spaces~\cite{DBLP:journals/tvcg/JerdingS98}.
Typically, we zoom and pan the space, causing objects of interest moving to off-screen space.
To overcome this trade-off between overview and detail, off-screen visualization techniques have been extensively researched. 
They are characterized by the idea of projecting off-screen located objects back to the available screen real estate. 
Detached visual cues overlay the visible space along the display edge and can encode spatial properties like direction, distance, up to full topology. 
The topology refers to spatial relations and properties unaffected by the change of shape or size of objects which is desirable to preserve~\cite{OxfordDictionaryofEnglish}. 

\begin{figure}[!ht]
	\centering
	\includegraphics[width=1.0\linewidth]{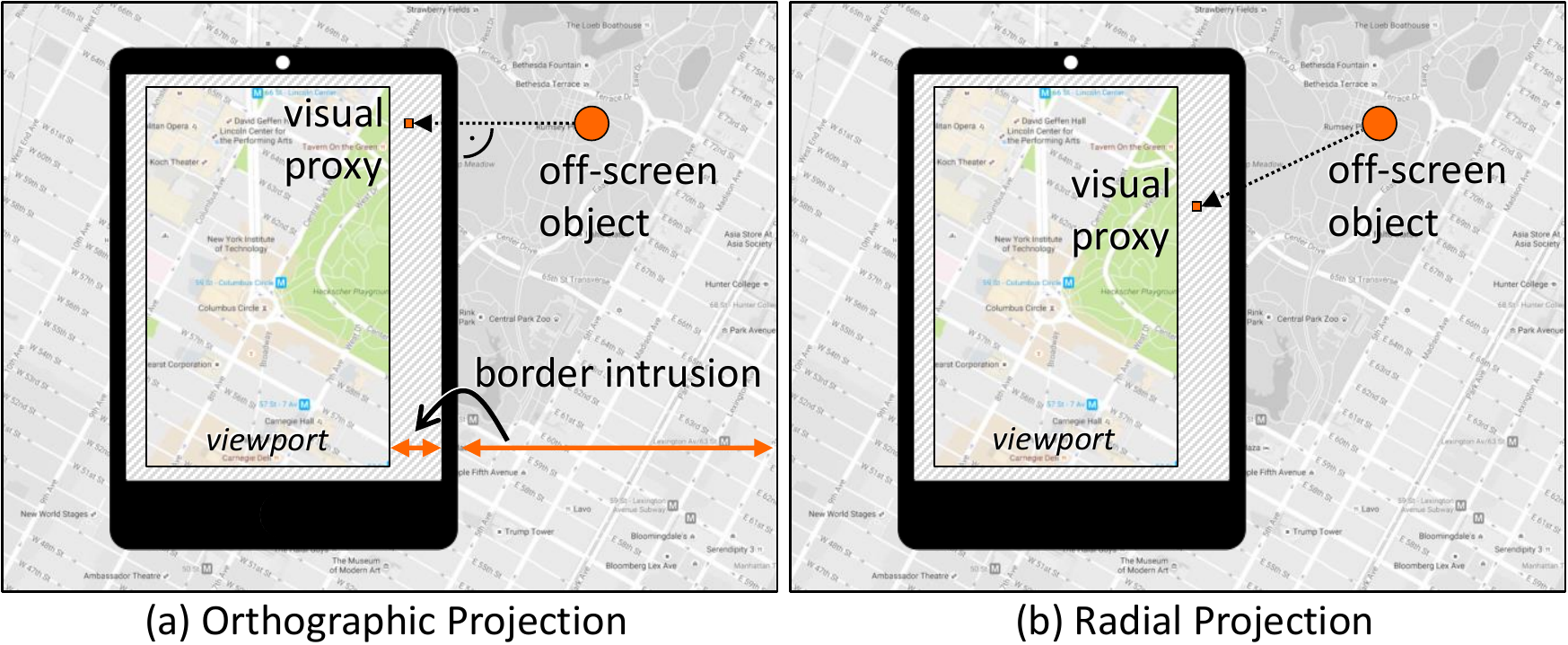}
	\caption{Off-screen projection strategies integrated in an adaptive intrusion border environment. The (a) orthographic strategy projects off-screen objects along a line perpendicular to the viewport. In contrast, (b) the radial strategy projects off-screen objects along a line towards a point of interest lying inside the viewport, which in our case is the center of the viewport. The intrusion of the border is adapted to the considered space in off-screen: The border's dimension on the right is relative to the space off-screen and thus the widest compared to the other sides.}
	\label{fig:radial_vs_ortho}
\end{figure}

Existing techniques \cite{DBLP:conf/chi/GustafsonI07,eurovisshort.20151125} encode the topology of off-screen objects in a dedicated border enclosing the visible space. 
The intrusion of the border is uniform on each side of the display. Distances between off-screen objects and the display are compressed proportionally into the border, allowing to bring off-screen objects efficiently in relation with each other.
However, due to the uniform border intrusion, the dimensions of the navigated space are not reflected.
To overcome this issue, we propose to adapt the intrusion of the border pursuant to the dimensions of its adjacent off-screen space, improving the awareness of the navigated space. 
Figure~\ref{fig:radial_vs_ortho} depicts an example: the border's intrusion on the right is proportional to the off-screen space on the right. 
The right border is the widest compared to the top, left, and bottom side. 
One problem arising is that off-screen points are also positioned proportional to the border size, which can be different on each side of the display.  
This requires to allocate the intrusion of the border to its actual size, namely the adjacent off-screen space, and only then being able to correctly relate the objects to each other. 

\looseness=-1
Another decision to take refers to the projection strategy, which indicates the direction to off-screen objects and enables efficient navigation. 
Existing techniques choose one out of two strategies to project off-screen objects back to the border: radial or orthographic. 
Both strategies are illustrated in Figure~\ref{fig:radial_vs_ortho}. 
Using the (a) orthographic projection strategy, off-screen objects are projected perpendicular to the viewport, whereas the (b) radial strategy projects off-screen objects along a line originating from the center of the viewport. 
The projection strategies have been evaluated for non-topology-preserving environments (e.g. \cite{DBLP:conf/nordichi/MullerLHB14}), however, are not applicable to topology-preserving environments using a border region. 
So far, no evidence is given for choosing one projection over the other in topology-preserving environments. 
\looseness=-1
In this paper, we conducted a controlled experiment to research the effect of the adaptive border intrusion to the data topology as well as the users' intuition regarding the projection strategy. 
The experiment consists of three consecutive tasks build upon a comprehensive derivation of the design. 
Our results indicate that an adaptive intrusion does not affect perceived relations of and between off-screen objects and that significantly more users apply the orthographic projection strategy. 
In preparation for our experiment, we reviewed existing off-screen techniques regarding encoded spatial characteristics, the usage of a border, and the projection strategy. 
Also, we conducted a pre-study using paper prototyping to confirm that the adaptive border design is understandable and it is ruled out to harm the experiment.
Based on the results of our experiment, we contribute design considerations regarding topology-preserving off-screen visualization. 
Furthermore, we discuss our results in consideration of the preservation of data characteristics. 

\section{Related Work}

To tackle the problem of providing overview, off-screen visualization techniques ``[...] modify how objects are rendered and can introduce proxies for objects that might not be expected to appear in the display at all'' \cite[p.~2:16]{DBLP:journals/csur/CockburnKB08}. 
This means, that off-screen located objects are projected back to the visible space using a different representation to emphasize certain characteristics of the data.
Off-screen techniques have been widely designed for navigation in networks or graphs \cite{DBLP:conf/chi/GeymayerSLSS14,DBLP:journals/cgf/GhaniRE11,DBLP:journals/cgf/MaySDK12,DBLP:conf/chi/MoscovichCHPF09}. 
Connected off-screen objects are projected to the screen real estate using along the line projection for efficient edge routing, which is comparable to the radial projection strategy except that it originates from multiple sources (the graph's nodes) within the viewport. 
An exception represents the approach by Frisch and Dachselt~\cite{DBLP:journals/ivs/FrischD13}, who discuss the projection strategy for navigating class diagrams. 
Objects in such diagrams are not connected via the shortest path, demanding a special solution. 
However, users expected a radial projection. 
We argue that it is not apparent which projection strategy to implement, explicitly for unconnected spatial objects in a topology-preserving environment. 
Table~\ref{tab:related_work} provides an overview of reviewed papers that respectively introduce a novel off-screen visualization technique based on unconnected objects. 
We discuss these papers focusing on the encoding (direction, distance, and topology), the usage of a dedicated border region, and the applied projection strategy.

\begin{table}[t]
	\centering
	\caption{Overview of surveyed papers that respectively introduce a novel off-screen visualization technique. Papers are ordered by year. Columns outline the relevant characteristics of each paper: the encoding of the technique (direction, distance, and topology), whether the visual cues are integrated into a border and the used projection strategy. The \protect\pie{360} indicates if a characteristic is fulfilled, the \protect\pie{180} if partly fulfilled, and the \protect\ding{51} whether a dedicated border is used.}
	\includegraphics[width=\linewidth]{rw_table_tr.tex}
	\label{tab:related_work}
	\vspace{-0.5cm}
\end{table}

\subsection{Encoding Direction, Distance, and Topology}
\looseness=-1
Each visual cue representing an off-screen object encodes at least one of the encodings: direction and distance. 
The direction indicates the route and the distance provides information on how far to pan in order to reach an off-screen object. 
While all papers in Table~\ref{tab:related_work} encode the direction, not all of them clearly indicate the distance (marked with a \pie{180}). 
Techniques using aggregation \cite{gonccalves2011halodot,eurova.20151103,eurovisshort.20151125} provide an approximation of the distance because aggregated points are visualized via a representative. 
The topology builds upon direction and distance and adds information on adjacent objects; the topology should show the object of interest as well as the spatial relations to other objects, either nearby or far off. 
Techniques like Bring\&Go (graph-based)~\cite{DBLP:conf/chi/MoscovichCHPF09} or Hopping~\cite{DBLP:conf/chi/IraniGY06} therefore use a radar-like representation and project off-screen located objects to the viewport in a radial manner. 
Other techniques such as EdgeRadar~\cite{DBLP:conf/chi/GustafsonI07} and Ambient Grids~\cite{eurovisshort.20151125} use a border region to compress the topological information of the respective adjacent off-screen space. 


\subsection{Dedicated Border Region}
\looseness=-1
According to Table \ref{tab:related_work}, the visualization of topology goes hand in hand with the application of a dedicated border region. 
In the first place, the pioneering work of Apperley et al. \cite{apperley_bifocal_1982} -- the bifocal display technique -- inspired the usage of a dedicated border region. 
The surrounding is distorted while the focus region is maximized. 
Applied to off-screen visualization, the off-screen space is compressed into the adjacent border, but in data-space and not in image-space. 
Besides the advantage of showing the data topology, the usage of a border also improves the awareness of empty areas. 
The border shows an explicit off-screen data distribution saving unnecessary panning operations to empty regions \cite{DBLP:conf/uist/JulF98}. 
Due to the uniform border size at each side of the display, present off-screen techniques \cite{DBLP:conf/chi/GustafsonI07,DBLP:conf/mhci/HossainHLI12,eurova.20151103,eurovisshort.20151125} fall short to reflect the off-screen space dimensions. 
As a result, one loses awareness of the position in space, particularly where the focus was set to.
For example, if one focuses the very left side of the data space, the border should indicate that only little data space remains on the left and that there is a bigger space to navigate on the right. 
To the best of our knowledge, the unawareness of the dimensions of the navigated space has yet not been addressed in the context of off-screen visualization. 
The solution we propose is to adapt the border intrusion on each side of the display individually. Figure \ref{fig:distance_scale} illustrates this concept: The image on the left shows the border intrusions in case the focus is positioned on the very left in data space. The right border intrusion is larger compared to the left border intrusion, because more space to navigate remains on the right hand side. 

\subsection{Projection Strategy}
The projection strategy is one of the most important design decisions to make; it determines the perceived direction and therefore affects the navigation route. 
CityLights \cite{DBLP:conf/chi/ZellwegerMGSB03} (non topology-preserving) and Ambient Grids \cite{eurovisshort.20151125} (topology-preserving) discuss the projections, but no evaluation, hence, no evidence is given for choosing one projection over the other in topology-preserving environments. M\"uller et al. \cite{DBLP:conf/nordichi/MullerLHB14} evaluated the projections for a non-topology-preserving environment using LEDs, which is why results are not applicable. In this paper, we make a first attempt to present empirical evidence for the right projection choice in topology-preserving off-screen environments.


\section{Design Space}\label{sec:designSpace}

Based on identified encodings of off-screen visualization techniques in Table~\ref{tab:related_work}, we discuss the design space of the experiment. This includes the visual abstraction of off-screen objects, the calculation of the border intrusion, and the projection. 
An adaptive border intrusion provides awareness of the data space dimensions and preserves the topology of off-screen objects. So far, no evidence has been given which projection strategy meets the users intuition in a topology-preserving environment.

\subsection{Visual Abstraction}
According to Cockburn et al.~\cite{DBLP:journals/csur/CockburnKB08}, off-screen visualization modifies the representation of off-screen objects. 
However, to indicate the presence of off-screen objects the visualization does not have to appear necessarily on-screen, as it is the case for techniques based on ambient light \cite{DBLP:conf/nordichi/MullerLHB14,DBLP:conf/chi/PertenederGLSH16}.
Apart from that, we distinguish between three classes of modifications: encoding of direction, distance, and data-specific characteristics. 
Various techniques exceed the encoding of direction via arrows and also encode distance, as depicted in Table~\ref{tab:related_work}. 
In addition, few techniques encode data-specific characteristics, such as a unique colored reference \cite{DBLP:conf/nordichi/MullerLHB14,DBLP:conf/chi/PertenederGLSH16}, the amount or density of clusters \cite{DBLP:conf/chi/GustafsonBGI08,eurova.20151103,eurovisshort.20151125}, among others. 

Beyond that, topology-preserving techniques reveal relations of off-screen objects to each other using direction and distance within a border. 
The representation of objects within the border has yet not exceeded simple rectangles~\cite{DBLP:conf/chi/GustafsonI07} and grid-based heatmaps~\cite{eurovisshort.20151125}. 
For our experiment, we use rectangles and apply a simple color coding to be able to distinguish between objects, but without a meaning attached. 

\subsection{Adaptive Border Intrusion} \label{sec:adative_border}

\begin{figure}[h]
	\centering
	\includegraphics[width=1.0\linewidth]{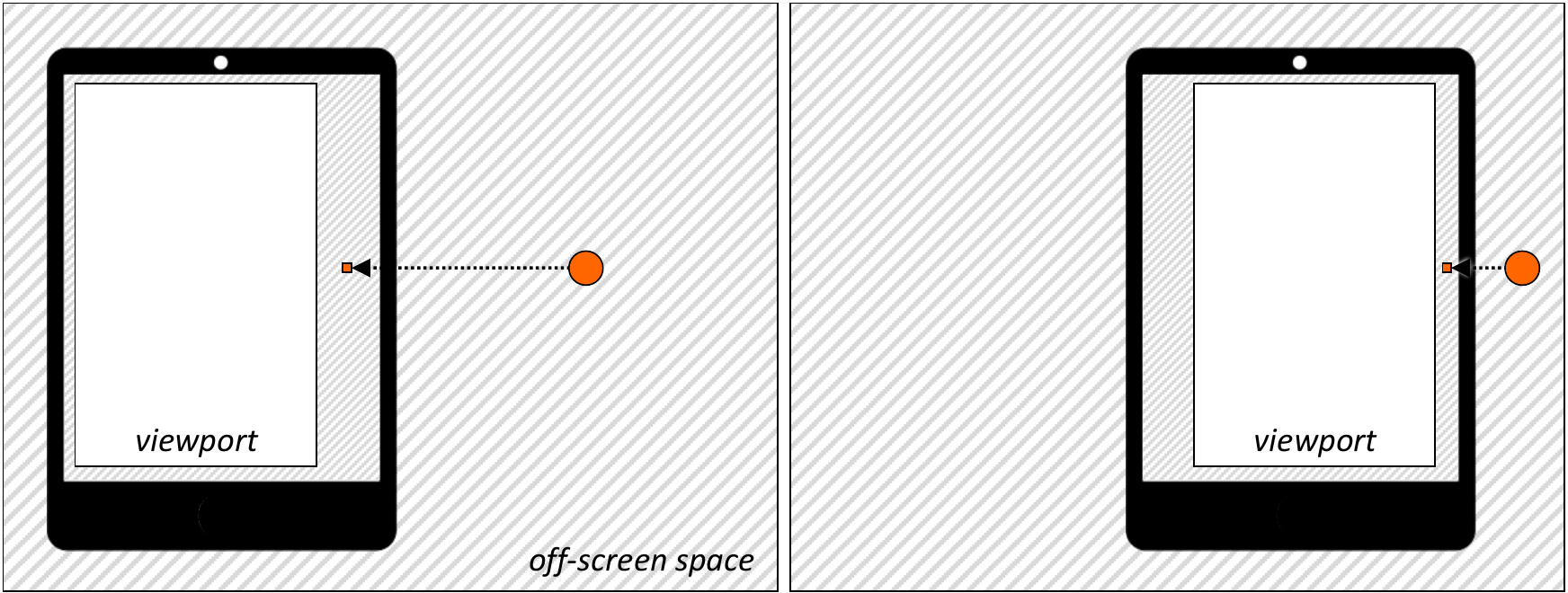}
	\caption{The two settings illustrate the effect of focusing different parts of the data space. Left: The focus is set on the very left in the data space. Right: The focus is set to the right in the data space. In each case, an off-screen point is positioned in the middle of the adjacent off-screen space. Depending on the focus position, the border adapts to the data space bounds. Independent of how big the off-screen space is, the centered off-screen point is projected to the respective middle of the border.}
	\label{fig:distance_scale}
\end{figure}

The border is placed along the display edge surrounding the viewport. 
In order to reflect the dimensions of the navigated space, we propose to adapt the intrusion of the border at each side of the display individually. 
This means, the border grows or shrinks proportionally to the adjacent off-screen space.
Furthermore, the border size needs to in- and decrease with the zoom level, as it is the case in multi-scale interfaces \cite{DBLP:conf/chi/FurnasB95}. 
Figure \ref{fig:distance_scale} illustrates the concept for two off-screen points positioned in the middle of the adjacent off-screen spaces. In both cases, the point is projected into the right border region. 
The adaptive border intrusion causes the border in the left image to be bigger compared to the border in the right image. Due to the adaptive intrusion, both off-screen points yet are projected into the center of the border. 
We dynamically calculate the intrusion of each side of the border while zooming and panning as follows: 
\begin{equation*}
size_x = \alpha \cdot \underbrace{\frac{zoom}{maxZoom}}_{1: \hspace{0.2em} zoom} \cdot \underbrace{min \left(1.0, \frac{d(vp, d\_bounds_x)}{d\_dimension_x} \right)}_{2: \hspace{0.2em} position \hspace{0.2em} in \hspace{0.2em} data \hspace{0.2em} space}
\end{equation*}

The dimensions of the border depend on two factors; (1) the zoom level, and (2) the distance to the outer bounds of the navigated space. 
$\alpha$ determines the maximum possible size of the border, which we set to $\alpha = 35$ pixels pursuant to EdgeRadar~\cite{DBLP:conf/chi/GustafsonI07} (we take up the impact of the border size in the discussion section). 
$\alpha$ is multiplied by the first factor, the zoom level: the bigger the zoom level, the more space is assigned to the border region. 
For the experiment, we set this factor to $1.0$, because we don't want the zoom level to affect the decisions of how off-screen objects relate to each other. 
The second factor determines the size of the border region according to the viewport position in the navigated space and is calculated for each side of the viewport $vp$ separately, which is why $x \in \{ top, left, bottom, right \}$. 
The restriction to $1.0$ ensures that the size of the border does not further increase if the navigated space is located entirely outside the viewport. 
Otherwise, the second factor is computed by dividing the distance of the viewport $vp$ to the outer bounds of the navigated space with the viewport dimension. For top and bottom, the $d\_dimension$ corresponds to the height of the viewport, the width otherwise. 

\subsubsection{Paper Prototyping}

\begin{figure}[h]
	\centering
	\includegraphics[width=1.0\linewidth]{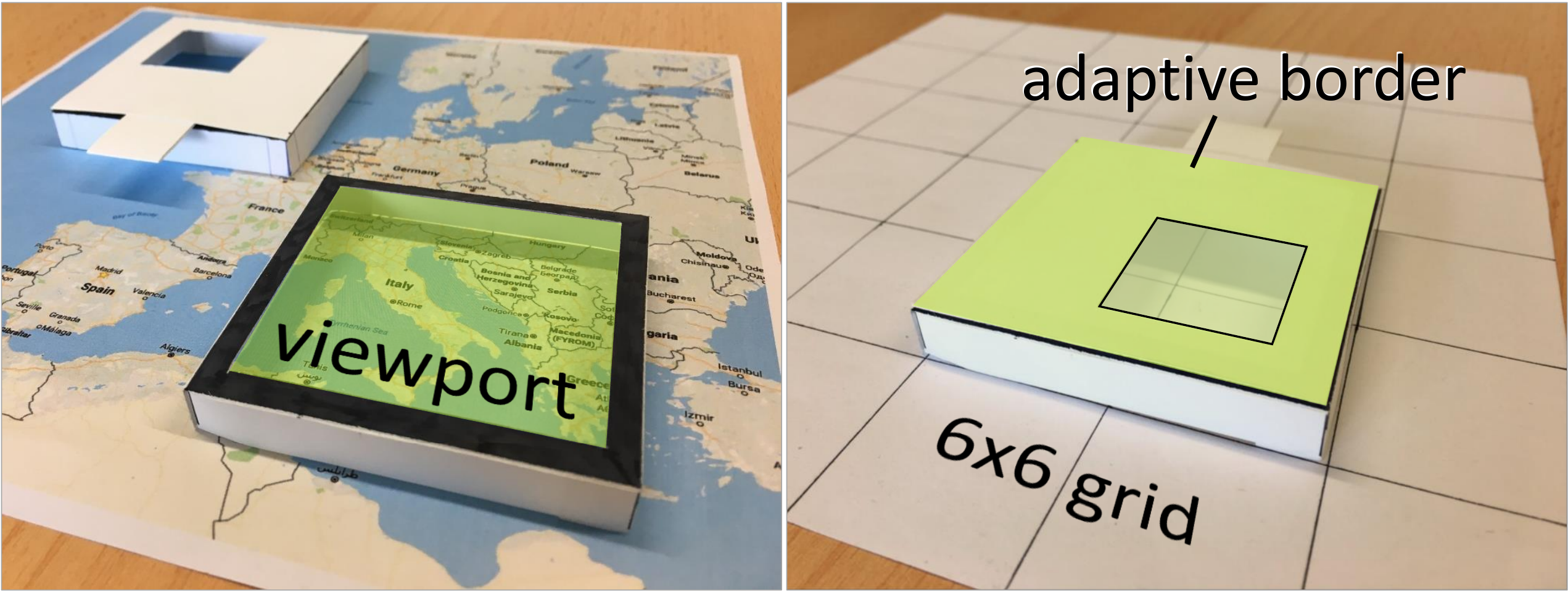}
	\caption{Paper prototype. Using adaptive border cutouts, any position of the viewport on the grid can be simulated.}
	\label{fig:paper_prototype}
\end{figure}

We carried out a preliminary study to investigate if the adaptive intrusion design of the border is well-received and does not mislead in any manner. 
Therefore, we created a paper prototype. 
We prepared a rectangular surface that consists of a grid of 36 equal sized cells (6x6). 
For the representation of the viewport, we built an elevated rectangle made of paper and cut out the inner area. 
The viewport covers four cells and can be freely placed on the grid surface similar to peephole interfaces. 
In addition, we set up five different border regions made of paper, which can be placed on the viewport. 
They are designed proportionally correct to the surface dimensions and allow to simulate any position of the viewport on the grid by rotating the border region: One for the center position. One for all positions where the viewport is moved horizontally or vertically by one cell. One for moving the viewport two cells, either horizontally or vertically. One for moving the viewport diagonally by one cell. One for moving the viewport diagonally by two cells. 

We recruited 5 participants (1 female) with normal or corrected-to-normal vision and did not report on color blindness. 
We tested with two elderly persons (69 and 73 years old) and 3 participants of age 25, 28, and 30. 
Among all participants, only one participant has worked with visualization as well as off-screen visualization before. 

\looseness=-1
Our setup is depicted in Figure~\ref{fig:paper_prototype}: We first introduced participants to our setup using a map. 
We placed the viewport on the map and demonstrated in one scenario how the intrusion of the border changes when navigating to the top left corner region. 
Then, we presented sequentially and in arbitrary order five different scenarios with no scenario being redundant. 
Participants were asked to place the viewport on the grid for each scenario separately. 
We recorded all actions on video. 

\looseness=-1
We measured the positioning error of the viewport in half and full grid cells. 
In total, participants placed the viewport with an average error of $0.12$ grid cells. 
Two out of the five participants did not perform any error. 
The two oldest participants mentioned that participating in this study ``gave them a reality check''. They used websites like Google Maps before, but did not fully understand the underlying principle. The paper prototype explained them in a simple manner how such applications operate. One interesting observation was that none of the participants asked for further explanations, yet well-performing the task. 
In consequence, we claim that the principle of an adaptive intrusion is easy to understand. 

\subsection{Projecting Off-screen Objects to the Border}

\looseness=-1
As introduced in the previous Section~\ref{sec:adative_border}, the adaptive border intrusion is calculated for each of the sides individually. The two main advantages are the preservation of the overall topology and the awareness of the surrounding data space dimensions. 
Yet, it has not been evaluated how the projection strategy affects the human perception of off-screen objects within a dedicated border region. 
Furthermore, it is unclear whether the adaptive border intrusion impairs the perception of off-screen relations and distances in any manner. 
In this study, we examined the two most prominent projection strategies: the orthographic and the radial projection strategy. 

\looseness=-1
One unanswered question refers to the choice of the distance encoding. 
The degree-of-interest function by Furnas~\cite{furnas_generalized_1986} proposes to consider nearby objects as more interesting than others. 
One can argue to adapt the distance function, especially in a topology-preserving environment, to this requirement. 
For example, a logarithmic-alike function grants nearby off-screen objects more space. 
While such a function increases the awareness of nearby objects, it interferes with the idea of putting objects in relation to each other plus deciding on the projection strategy. 
The primary goal of this study is to investigate the effect of the adaptive border and the projection strategy, which is why we choose a linear distance encoding.

\section{Experiment}\label{sec:experiment}

\begin{figure*}[h]
	\centering
	\includegraphics[width=0.98\linewidth]{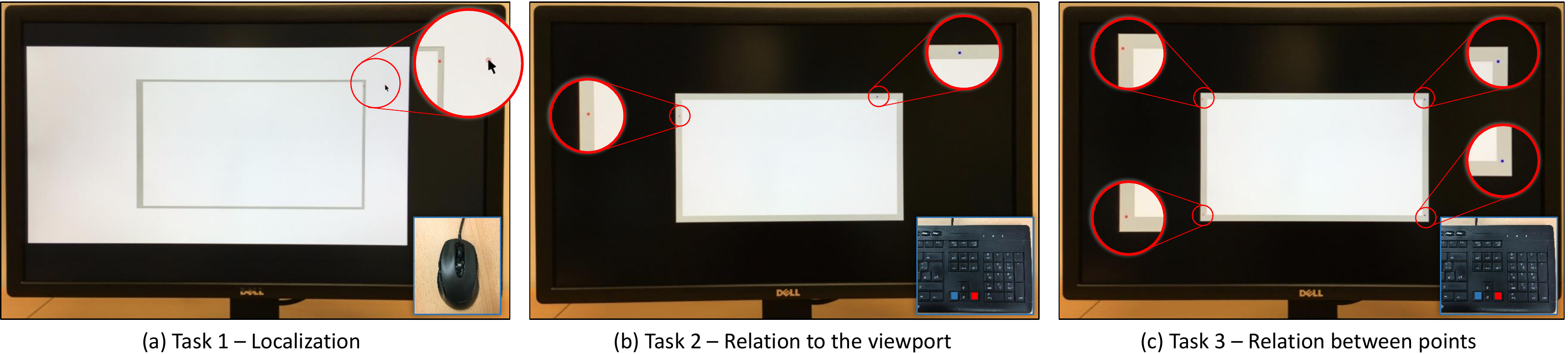}
	\caption{Taken pictures of all three tasks. (a) Task 1: one point was positioned in the top right border region and the user clicks in the pointing hub (slightly shifted to the right) where she assumed the position. (b) Task 2: one red point was placed on the top left and one blue point was placed on the top right. (c) Task 3: two red points were positioned on the left and two blue points were positioned on the right of the viewport.}
	\label{fig:setup}
\end{figure*}

We conducted a controlled experiment to investigate two effects. First, we investigated the effect of the adaptive border intrusion, in particular, whether the adaptive intrusion design affects the perception compared to the uniform intrusion design. Second, we examined how users perceive off-screen objects that are projected back to the screen real estate, namely the projection strategy. The projection strategy has not been evaluated in the context of a topology-preserving border region so far. 
All tasks of the experiment were carried out using the same apparatus. 

For the description of this experiment, we modify our terminology as follows: the data space refers to the space containing the data. The data space is the space to be navigated. 
Furthermore, off-screen objects are denoted as off-screen points. 
This is because we aim at investigating perceived spatial information and not any other characteristics, which may be assumed when using the term \textit{object}. 

\subsection{Tasks}
We reviewed existing off-screen techniques (see Table~\ref{tab:related_work}) regarding the evaluation type and tasks. 
Based on the related work, we derived three consecutive tasks for our experiment: 
First, we tested how users localize the position of off-screen points in the data space. 
Second, we tested how users judge relations of off-screen points to the viewport and third, we tested how users judge relations between off-screen points. 
Following, we describe the three tasks. 

\subsubsection{Task 1: Localization of Off-screen Points}
The first task aimed to investigate the users' intuition regarding the projection strategy. 
We tested for the orthographic and the radial projection strategy. 

\begin{figure}[h]
	\centering
	\includegraphics[width=1.0\linewidth]{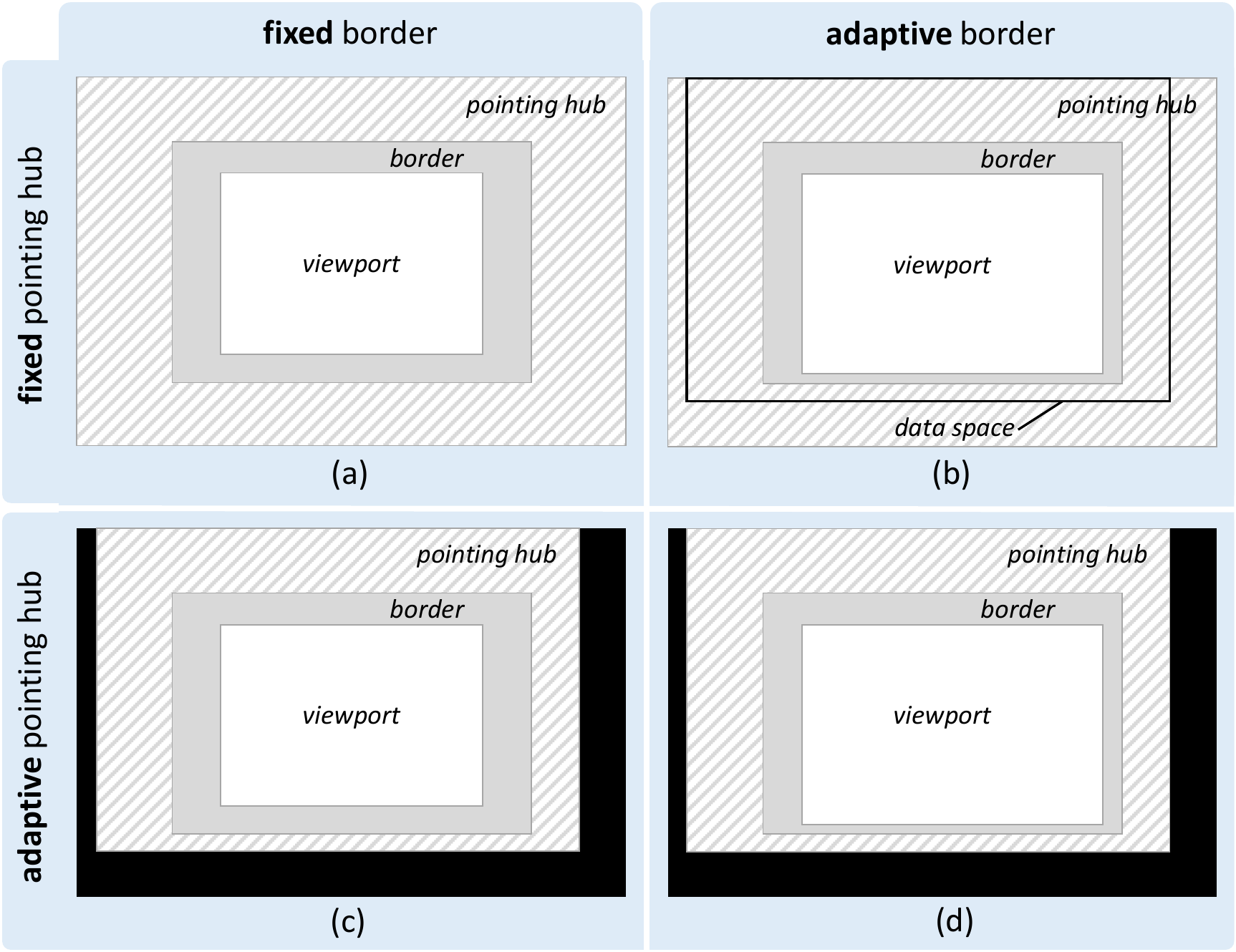}
	\caption{Differentiation of cases for Task 1. Four combinations of fixed border, adaptive border, fixed pointing hub, and adaptive pointing hub were presented to the participant to test the projection strategy and the adaptive border design.}
	\label{fig:pointing_hub_design}
\end{figure}

\looseness=-1
Based on the idea of Sparkle~\cite{DBLP:conf/nordichi/MullerLHB14}, the display was divided into two areas: 
The center area imitated the viewport including the off-screen visualization by means of a dedicated border. The surrounding area represented the maximum possible data space which did not exceed the display bounds.
We successively presented a projected point in the border region and asked participants to click in the off-screen space where they expected the initial position of the data point. 

The off-screen area that allows participants to mark the assumed position is called \emph{pointing hub}. 
We outline all presented cases in Figure~\ref{fig:pointing_hub_design}.
First, we distinguish between the fixed and the adaptive border design, depicted left and right, respectively. 
The fixed border design corresponds to a uniform border that does not adapt to the data space dimensions, whereas the adaptive border design reflects the dimensions. 
Then, we distinguish between a fixed and an adapted pointing hub. 
A fixed pointing hub (top row in Figure~\ref{fig:pointing_hub_design}) stands for a fully extended area surrounding the off-screen visualization. An adapted hub (bottom row in Figure~\ref{fig:pointing_hub_design}) stands for a variable extent surrounding the off-screen visualization. 
Interaction is only allowed within the crosshatched pointing hub. 
We distinguish between a total of four cases. 
Figure~\ref{fig:pointing_hub_design} (a) depicts the case of the fixed border design and a full pointing hub extent. 
Here, the off-screen space dimensions correspond to the pointing hub dimensions. 
Figure~\ref{fig:pointing_hub_design} (b) shows the same case, but for an adaptive border design. The pointing hub is also fully extended, however, the off-screen space dimensions do not correspond to the pointing hub dimensions; participants were asked to count back the presented adaptive border. 
The bottom row in Figure~\ref{fig:pointing_hub_design} depicts the cases for an adaptive pointing hub. 
This means, the pointing hub is adapted to the dimensions of the off-screen space. 
Figure~\ref{fig:pointing_hub_design} (c) shows the fixed border design, however, the off-screen space dimensions and thus the pointing hub dimensions, are not fully extended. 
Figure~\ref{fig:pointing_hub_design} (d) shows the same off-screen space design, but the border is adapted to the surrounding off-screen space. 

\looseness=-1
Figure~\ref{fig:setup} (a) depicts the setup and one example: In this example, a red point is placed on the top right and the participant uses the mouse to click on the assumed position, which is slightly shifted to the right. The depicted case corresponds to the adaptive border design combined with a variable off-screen space and positioning hub (compare to Figure \ref{fig:pointing_hub_design} (d)).  

\subsubsection{Task 2: Relation to the Viewport}
In this second task, we intended to investigate the influence of the adaptive border design to relations between off-screen points and the viewport. 
Therefore, we tested the fixed border design versus the adaptive border design. 

\looseness=-1
We successively presented to participants an off-screen visualization that contained two projected points, one was filled blue and the other one red. 
Participants were then asked to decide which projected off-screen point was farther away from the viewport. 
Figure~\ref{fig:setup} (b) depicts the setup and an example: In the example, one red point is placed left and one blue point at the top right. 
Two keys on the keyboard were marked in the respective colors, so that participants could efficiently decide on the distances. 
In contrast to Task 1, the surrounding area of the off-screen visualization was colored in black. 
However, the full extent to the display size still corresponded to the possibly full extent of the off-screen space, providing participants an area of reference when counting back the border intrusion as well as the projection.

\subsubsection{Task 3: Relation between Points}
This task tested relations between points integrated into the adaptive border design. 
We tested the fixed border design against the adaptive border design.

Successively, two blue and two red projected off-screen points were presented to the participant, who had to decide which points are closer to each other: the blue ones or the red ones. 
Two keys were marked in the corresponding colors on the keyboard, as depicted in Figure~\ref{fig:setup} (c). 
The Figure illustrates an example, were two red points are located on the left and two blue points on the right. 
As described in Task 2, the surrounding area of the off-screen visualization was colored in black and the full extent to the display size reflected the possibly maximal dimensions of the off-screen space.

\subsection{Data Generation}
In this section we focus on the generation of data points for the described tasks. 
Therefore, we first derived cases that show the biggest difference between projection strategies and then generated the data points. 
We chose cases showing the biggest difference, because we aimed at confronting participants with the biggest possible discrepancy between the orthographic and radial projection strategy. This way, we can make a clear statement and provide guidelines which projection strategy to use in a topology-preserving environment. 
The biggest difference can be defined as the maximum Euclidean distance between projected points, which represent the same off-screen point, but use a different projection strategy. 
Consider a point $p$ located in off-screen space. 
The orthographic strategy projects point $p$ to a location $p_{O}$ and the radial strategy projects the point to a location $p_{R}$ within the border. 
The aim was to identify cases, for which the distance between $p_{O}$ and $p_{R}$ is maximal. 
This way, results are interpretable without being biased by the intended projection strategy. 

Following, we first derive aforementioned extreme cases, and then derive the data generation on a per-task basis. 

\begin{figure}[b]
	\centering
	\includegraphics[width=1.0\linewidth]{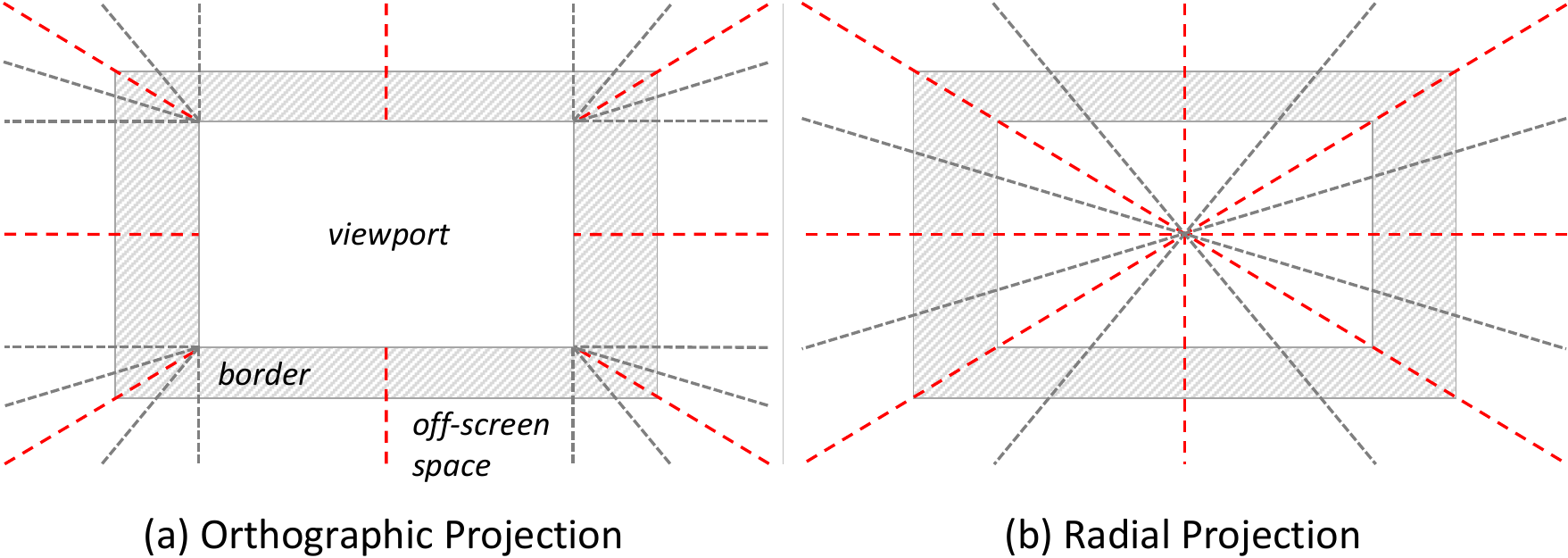}
	\caption{\looseness=-1 Distinction of projection strategies with respect to identical results. The dotted lines highlight differences and commonalities between projection strategies. (a) shows the orthographic strategy: off-screen points are projected perpendicular to the viewport. In the corner areas, points are projected evenly in $x$- and $y$-direction. Using the (b) radial projection strategy, off-screen points are projected along the line towards the center of the viewport. The red colored lines represent the projection lines along which the results of the orthographic and radial projection are identical.}
	\label{fig:projection_same_different}
\end{figure}

\subsubsection{Derivation of Extreme Cases}
A good positioning of points minimizes the randomization of locations. 
Furthermore, it makes results interpretable, because results are not in favor of a specific projection strategy; this means, the presented projected point does not pretend a specific projection strategy. 
We derived various areas in off-screen space, which meet our requirement of providing the biggest difference by means of the projection result. 
We call these: \emph{extreme cases}. 
To derive the extreme cases, we first focused on the commonalities between both projection strategies. 

We sketched both projection strategies in Figure~\ref{fig:projection_same_different}. 
The dotted lines illustrate the concepts of how corresponding projection strategies project off-screen located points back to the screen real estate. 
On the left hand side, the (a) orthographic projection strategy is characterized by the uniform projection perpendicular to the viewport, either in $x$-direction for left/right sides or in $y$-direction for top/bottom sides. 
The corner areas call for a special treatment, which is an even projection in $x$- and $y$-direction similar to the bifocal display technique~\cite{apperley_bifocal_1982}. 
This means, off-screen points located in the corner areas are projected, similar to the radial strategy, towards the respective corner of the viewport. 
For the (b) radial projection strategy, depicted on the right hand side, off-screen points are projected along the line towards the center of the viewport. 
The red colored lines represent the projection lines, along which both projection strategies yield equal results. 
This is, all off-screen points placed along the red lines produce the same result when being projected, either orthographically or radially. 

For the remainder of this section, we reference to all positions along the red diagonal lines as function $f_{d}$ and along the red median lines as function $f_{o}$.
The functions $f_{d}$ and $f_{o}$ define all points which are identically projected to the border, whether following the orthographic or the radial projection strategy. 
We assume that the extreme cases lie in the area enclosed by the functions $f_{d}$ and $f_{o}$. 

\begin{figure}[t]
	\centering
	\includegraphics[width=1.0\linewidth]{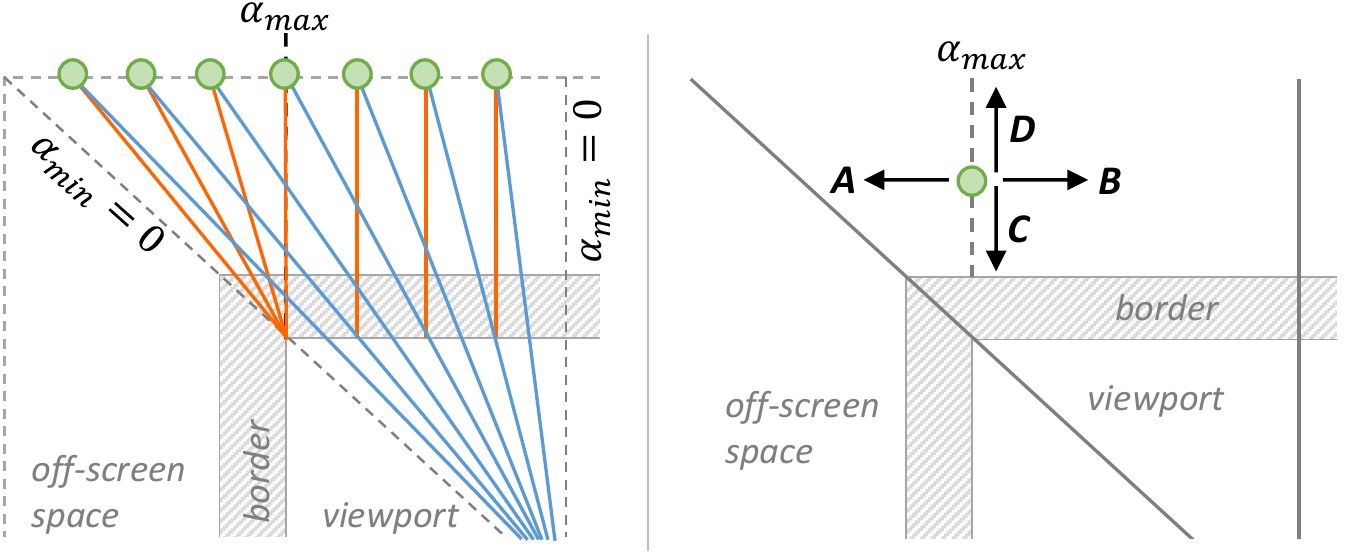}
	\caption{Left: Explanation of the difference between projection strategies by inspecting the angle $\alpha$. $\alpha$ describes the angle between lines, along which the \textbf{point} (green) is projected, either \textbf{orthographically} (orange lines) or \textbf{radially} (blue lines). Right: Degrees of freedom an off-screen point can be moved. $\alpha_{max}$ is defined along the line, which represents the transition between side and corner area. Along this line, $\alpha_{max}$ determines that the radial and orthographic projection strategy differ the most.}
	\label{fig:projection_discussion_positioning}
\end{figure}

In order to inspect the behavior of projections in this area, we investigated the angle $\alpha$ between projection strategies. 
Consider one point being projected to the border region. 
Using the orthographic strategy, the point is projected along a line perpendicular to the viewport. 
Using the radial strategy, the point is projected along a line towards the center of the viewport. 
The angle $\alpha$ is the angle between these two projection lines. 
Figure~\ref{fig:projection_discussion_positioning} illustrates an example, based on which we can visually derive the extreme cases. 
The functions $f_{d}$ and $f_{o}$ define all cases, in which projections are identical; i.e. $\alpha=0$.
Figure~\ref{fig:projection_discussion_positioning} depicts on the left hand side what happens to $\alpha$ when moving the off-screen point (colored in green) between the median $f_{o}$ and the diagonal line $f_{d}$. 
For example, when moving from the median towards the diagonal line, $\alpha$ first increases and after having reached its maximum value $\alpha_{max}$, it decreases again. 
For a quadratic viewport, $\alpha_{max}$ is reached at the transition between side and corner area. 
For a non-quadratic viewport $\alpha_{max}$ moves towards $f_{d}$ or $f_{o}$, respectively. 

Based on the data point being placed on the transition between side and corner area, we distinguished between four cases how $\alpha$ changes when moving the point. 
Figure~\ref{fig:projection_discussion_positioning} illustrates these cases on the right hand side: 
Moving towards \textbf{A} or \textbf{B}, $\alpha$ reaches the value $0$, as already discussed. 
Moving the point towards \textbf{C}, $\alpha \rightarrow \frac{1}{2} \pi$. 
For \textbf{D}, $\alpha \rightarrow 0$, which also decreases the projection error. 
For the sake of simplicity, we assume a quadratic viewport for the data generation, because the shift of $\alpha_{max}$ does not drastically affect the projection result. 
We call the line, along which $\alpha_{max}$ is placed, in the following: axis. Because of symmetry, there are eight extreme case axes, on which we positioned the data points. 



\subsubsection{Data for Task 1}
We presented the design of the first task in Figure~\ref{fig:pointing_hub_design}. 
For cases (b), (c), and (d), in which the off-screen space is not fully extended, we chose the extent on a per-side basis. 
One side was fully extended in order to provide participants an area of reference. 
For the remaining three sides, we randomly chose the extent to be within the inner 50\% of the possibly full extent. 
To choose the extent within the inner 50\% implies that we do not test for border cases, which refer to special cases such as positioning the data point just right at the transition between off-screen space and viewport, or at the very outer bounds. Therefore, we ensured a fair distribution of points. 
The allocation of all sides was randomly chosen. 
The presented data points were positioned along the 8 axes, which were identified as extreme cases. 

\subsubsection{Data for Task 2}
\begin{figure}[h]
	\centering
	\includegraphics[width=1\linewidth]{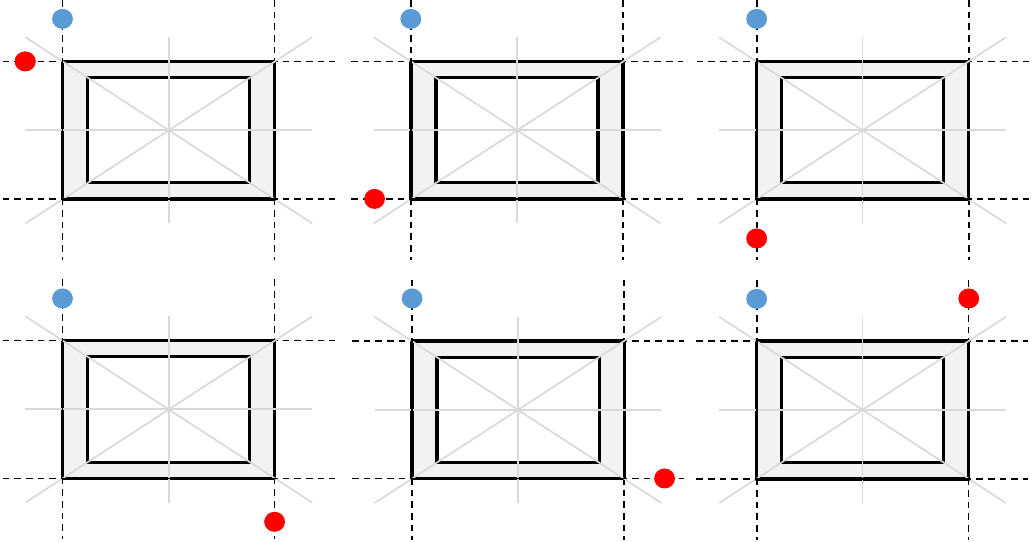}
	\caption{Tested cases of the second task. Due to the requirement that two points shall not be placed on the same axis, and symmetry of cases, six cases remain. Each case can be randomized by means of rotation.}
	\label{fig:task2_cases}
\end{figure}

In the second task, we presented two points at a time. 
One point was filled blue, the other point was filled red. 
For the generation of the data space, same as in Task 1, we chose one side to be fully extended as area of reference. 
For the remaining three sides, we chose the extent to be within the inner 50\% of the possibly full extent. 
Data points were positioned within the extent of derived axes that describe the extreme cases in data space, respectively. 
Overall, there exist eight axes (two on each side of the viewport) along we can place the points. 
In order not to allow comparing distances between points being placed on the very same axes, we only used cases where points were placed on different axes. 
Also, we considered the symmetry of cases. 
Figure~\ref{fig:task2_cases} depicts all considered cases. 
In total, there exist six cases.

\subsubsection{Data for Task 3}
In the third task, four points were presented to participants at a time. 
Two points were filled blue and two points were filled red. 
For the generation of the data space, same as in Task 1, we chose one side to be fully extended as area of reference and the remaining sides to be extended within the inner 50\% of the possibly full extent. 
Data points were positioned within the extent of derived axes that describe the extreme cases in data space, respectively. 
Based on the same requirements of Task 2, that red and blue off-screen points shall not be placed on the same axis and having regard to symmetry, we derived 36 cases. 
Out of this 36 cases, we randomly picked six cases for each participant but made sure that all cases were evenly covered among all participants.

\subsection{Hypotheses}
Considering the carried out paper prototyping evaluation as well as the systematic evaluation of projection strategies, we derive the following hypotheses on a per task basis:

\subsubsection{Localization of Off-screen Points}
\begin{description}
	\item[H1.1:] \textit{The orthographic projection will be chosen over the radial projection strategy.} Given related work and our conducted systematic evaluation, the orthographic projection is preferred over the radial projection by participants. 
	
	\item[H1.2] \textit{The task performance time will be higher with the fixed border design than with the adaptive border design.} In cases, in which participants need to count back the off-screen space dimensions, we expect participants to be slower. 
\end{description}

\subsubsection{Relation of Off-screen Points to the Viewport}
\begin{description}
	\item[H2.1:] \textit{The adaptive intrusion will not negatively influence the perception of relations of off-screen points.} The task performance accuracy using the adaptive border design will be comparable to the fixed border design. The adaptive border does not provide an enhanced perception of off-screen objects, but an overview of the data space dimensions.  
	
	\item[H2.2:] \textit{The task performance time will be higher for the fixed border design than for the adaptive border design.} This is due to the additional step of counting back the off-screen space for the adaptive border design.
\end{description}

\subsubsection{Relation of Off-screen points to Each Other}
\begin{description}
	\item[H3.1:]  \textit{The adaptive intrusion will not negatively influence the perception of relations between off-screen points.} The task performance accuracy using the adaptive border design will be comparable to the fixed border design. The adaptive border does not provide an enhanced perception of off-screen objects, as described in \textbf{H2.1}.
	
	\item[H3.2:] \textit{The task performance time will be higher for the fixed border design than for the adaptive border design.} This is due to the additional step of counting back the off-screen space for the adaptive border design.
\end{description}

\subsection{Experimental Design}
\looseness=-1
This study used a within-subject design. 
The independent variables were for Task 1: border and pointing hub. 
For Task 2 and Task 3, the independent variable was: border. 
There were three dependent variables for Task 1: error distance, time, and projection strategy. 
For Task 2 and Task 3, the dependent variables were accuracy and time. 
In the following section we discuss the derivation of the data and thus the trials.

\subsection{Participants}
We recruited 18 participants (2 female) mainly from the local student population. 
All participants had normal or corrected to normal vision. 
The age ranged from 21 to 68 years (Median age 28) with 3 participants reporting previous experience with off-screen visualizations. 
However, all participants were familiar with basic visualization techniques (i.e. line and pie charts, among others). 

\subsection{Apparatus}
\looseness=-1
The studies were conducted using a 27" monitor, one QWERTY keyboard, as well as a cord mouse. 
The display has a resolution of 2580x1440 pixels and was divided into two areas. 
The off-screen visualization was positioned in the middle of the display with the dimensions 1920x1080 simulating common 24" screens. 
The surrounding area represented the maximum possible data space and was communicated as this to participants. 

\subsection{Procedure}
The experiment was carried out in a quiet room at our university. 
Each participant was placed in front of the monitor and received an introduction to the topic of off-screen visualization using examples. 
In order not to prime participants for a specific projection strategy, we did not mention the projection strategy at all and only explained examples which are same for the orthographic as well as radial projection strategy. 
Also, we provided a comprehensive explanation of the adaptive border intrusion. 
During the study, the experimenter and the participant were the only persons present. 

The tasks were ordered from easy to difficult starting with Task 1, then Task 2, and finally Task 3. 
Before each task, the experimenter explained the task as well as necessary interactions. 
For each task, the participant stepped through a short training session using the default projection cases that did not suggest one of the two projection strategies. 
After each task, the participant had a short break. 
Following, we provide an overview of the amount of performed trials on a per task basis. 
As aforementioned, experimental factors were randomized and did not follow any defined order. 

For Task 1, we collected the task performance time as well as the distances between the marked position and the retraced orthographically and radially projected case. This means, we can argue if the participants assumed the position of the initial point to be nearer the orthographic or the radial projection. 
\begin{center}
	\footnotesize
	\begin{tabular}{rll}
		2 & border properties (\textit{F}, \textit{A}) & \texttimes \\
		2 & pointing hub properties (\textit{F}, \textit{A}) & \texttimes \\
		8 & repetitions  & =\\
		\hline
		32 & trials per participant & \texttimes \\
		18 & participants & = \\
		\hline
		\textbf{576} & \textbf{trials in total for Task 1} \\
	\end{tabular}
\end{center}


For Task 2 and 3, we collected the task performance accuracy and time. 
For all tasks, the overall order of trials was randomized. 
\begin{center}
	\footnotesize
	\begin{tabular}{rll}
		2 & border properties (\textit{F}, \textit{A}) & \texttimes \\
		12 & repetitions  & =\\
		\hline
		24 & trials per participant & \texttimes \\
		18 & participants & = \\
		\hline
		\textbf{432} & \textbf{trials} & \texttimes \\
		2 & tasks (\textbf{T2}, \textbf{T3}) & =\\
		\hline
		\textbf{864} & \textbf{trials in total for Task 2 and Task 3} \\
	\end{tabular}
\end{center}

\section{Results}
We only report statistically significant results ($p < .05$) from our quantitative analysis and refer to qualitative feedback in the discussion section.

\subsection{Task 1 - Localization}



Because of the non-normal nature of our data we used a non-parametric Friedman's test to compare the error distance and participants' projection strategy. For analyzing the completion time, we applied a two-way repeated measures ANOVA.

\begin{figure}[h]
	\centering
	\includegraphics[width=0.95\linewidth]{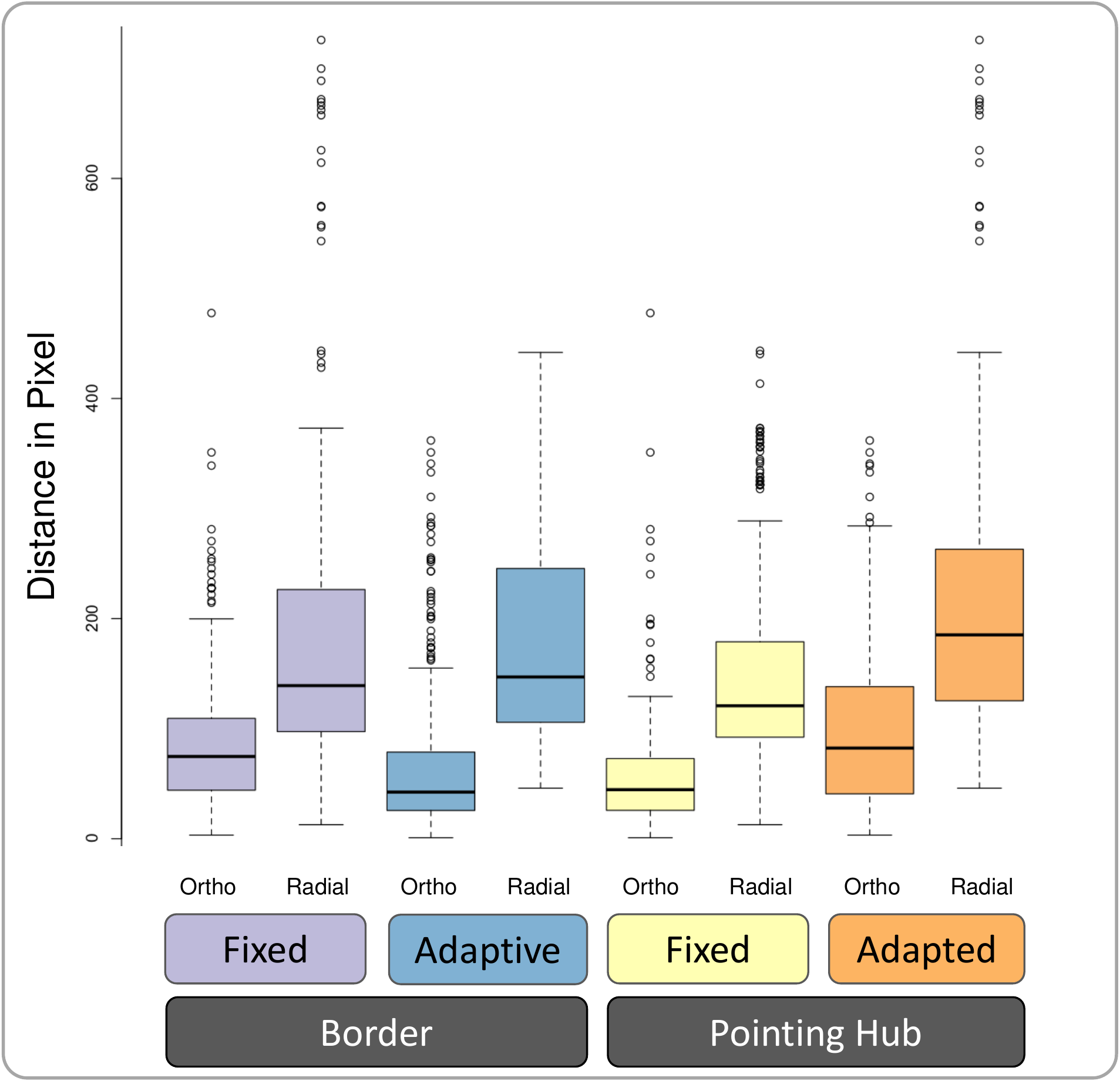}
	\caption{Error Distance: the box plots represent the Euclidean distance (in pixels) between the selected position and the location of both projections (orthographic and radial), respectively. Independent of the design, participants tend to project data points orthographically.}
	\label{fig:task1_results}
\end{figure}

\begin{figure*}[h]
	\centering
	\includegraphics[width=1.0\linewidth]{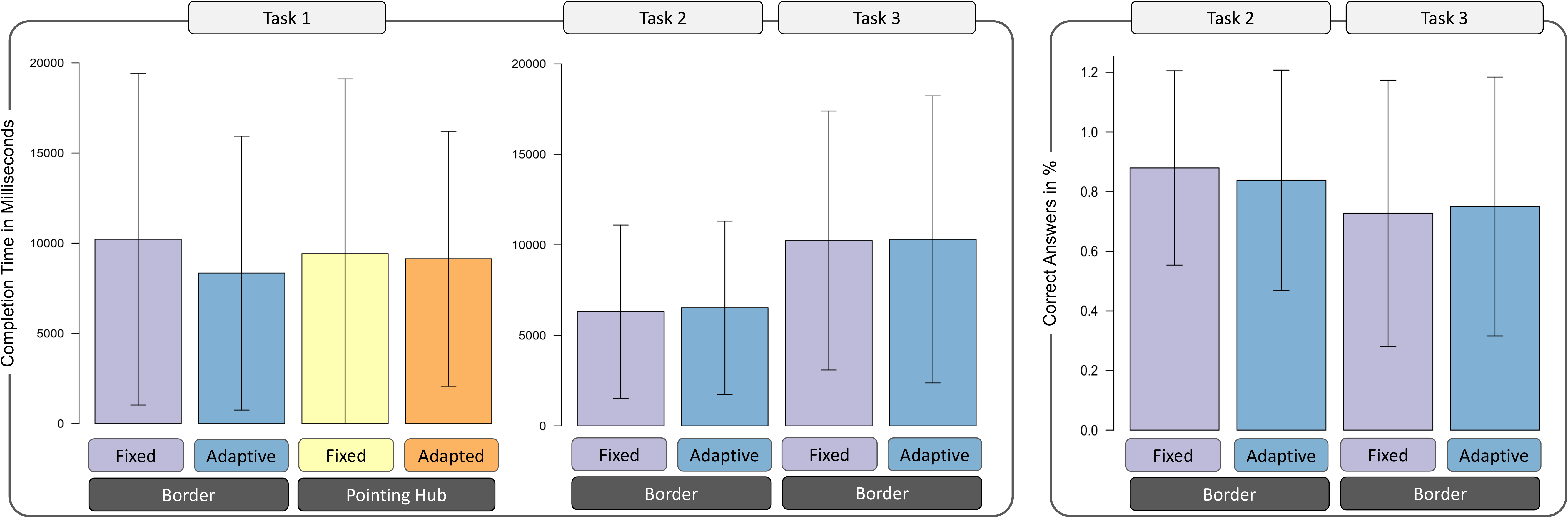}
	\caption{Completion Time and Accuracy: Bar charts with mean and standard deviation showing the completion time in milliseconds for all three tasks and the percentage of correct answers for task 2 and 3. The adaptive border has a positive effect on the completion time for Task 1 and does not perform significantly worse for the other two tasks.}
\end{figure*}

\subsubsection{Error Distance}
There was an overall significant effect of \emph{projection strategy} on \emph{error distance} ($\chi^2(1, N = 576) = 18, p < .001$).\\
Post-hoc tests revealed that error distance was significantly lower for the orthographic projection strategy  ($78.3 px$) compared to the radial projection strategy ($183.4px,p<.001$).

\textbf{Border:} There was an overall significant effect of \emph{border} on \emph{error distance} for both the orthographic projection strategy ($\chi^2(1, N = 288) = 5.6, p < .05$) and the radial projection strategy ($\chi^2(1, N = 288) = 18, p < .001$).\\
Post-hoc tests revealed that error distance was significantly lower for the adaptive border (orthographic: $69.5 px$, radial: $177.9 px$) compared to the fixed border (orthographic: $87.0 px$, radial: $188.9 px;p<.05$).

\textbf{Pointing Hub:} There was an overall significant effect of \emph{pointing hub} on \emph{error distance} for both the orthographic projection strategy ($\chi^2(1, N = 288) = 18, p < .001$) and the radial projection strategy ($\chi^2(1, N = 288) = 18, p < .001$).\\
Post-hoc tests revealed that error distance was significantly lower for the fixed pointing hub (orthographic: $56.1 px$, radial: $152 px$) compared to the adapted pointing hub (orthographic: $100.4 px$, radial: $214.8 px;p<.001$).

\subsubsection{Projection Strategy}
Figure~\ref{fig:task1_results} illustrates high level results. Overall, participants preferred an orthographic projection strategy ($94.4\%$).


\textbf{Border:} There was a significant effect of \emph{border} on \emph{projection strategy} ($\chi^2(1, N = 288) = 7.36, p < .01$).\\
Post-hoc tests revealed that participants used more often an orthographic projection strategy when working with the adaptive border ($98.6\%$) compared to the fixed border ($90.3\%,p<.005$).

\textbf{Pointing Hub:} There was a significant effect of \emph{pointing hub} on \emph{projection strategy} ($\chi^2(1, N = 288) = 4.5, p < .05$).\\
Post-hoc tests revealed that participants used more often an orthographic projection strategy when working with the adapted pointing hub ($96.2\%$) compared to the fixed pointing hub ($92.7\%,p<.05$).

\subsubsection{Completion Time}
There was an overall effect of \emph{border} on \emph{completion time} $(F_{1,17}=9.3, p<.01)$.\\
Post-hoc tests revealed that participants were faster when working with the adaptive border ($8.3sec$) compared to the fixed border ($10.2,p<.01$).

\subsection{Task 2 - Relation to the Viewport}

\subsubsection{Error Rate}
No significant results can be reported. Participants tend to answer equally correct for both the fixed border ($88\%$) and the adaptive border ($83.8$).

\subsubsection{Completion Time}
There is no significant effect between \emph{border} and \emph{completion time}. Participants were slightly faster when working with the fixed border ($6.3 sec$) compared to the adaptive border ($6.5 sec$)

\subsection{Task 3 - Relation Between Points}

\subsubsection{Error Rate}
No significant results can be reported. Participants tend to answer equally correct for both the adaptive border ($75\%$) and the fixed border ($72.7$).

\subsubsection{Completion Time}
There is no significant effect between \emph{border} and \emph{completion time}. Participants were slightly faster when working with the fixed border ($10.2 sec$) compared to the adaptive border ($10.3 sec$)

\section{Discussion}


Following, we discuss the result of our experiment as well as the individual feedback and derive Design Considerations (DC). 
Overall, the adaptive border design was well received as already assumed by our preliminary study using paper prototyping. Furthermore, we can show that even in a topology-preserving off-screen environment users tend to apply the orthographic projection strategy. 

\subsection{Projection Strategy}
Our results indicate that participants selected an orthographic projection strategy significantly more often confirming H1.1. 
The error distance was significantly lower for the orthographic projection strategy than for the radial projection strategy. 
Additionally, 17 out of 18 participants reported that the orthographic projection strategy seems more intuitive. 
In consideration of related work (see Table~\ref{tab:related_work}), we can give evidence that the orthographic strategy is preferred over the radial strategy for topology-preserving off-screen visualization. 
This finding is of particular interest compared to state-of-the-art off-screen techniques, which explicitly do not apply a border, because points that are integrated into a border do not preset a certain direction. 
One example represents Wedge~\cite{DBLP:conf/chi/GustafsonBGI08}, which encodes the direction already within the visual cue. 

Further intriguing feedback was collected from one participant, who argued that even though the orthographic projection strategy seems intuitive, he would have used the radial projection strategy for implementation. This means, the participant finds the radial strategy correct from an implementation point of view. We can learn from his feedback to first focus on the consumers and their needs before creating a solution. \\



\noindent \emph{\textbf{DC1:} The orthographic projection strategy in the best choice for topology-preserving off-screen visualizations of unconnected point data.} 

\subsection{Adaptive vs. Fixed Border}

The results indicate that the adaptive border design does not negatively influence the perception of relations of and between off-screen points, confirming H2.1 and H3.1.
Furthermore, the results of Task 1 indicate that participants were significantly more accurate counting back the position of projected off-screen points using the adaptive border design. 
However, participants were not significantly faster with the fixed border design, thus rejecting H1.2, H2.2, and H3.2. In other words, the task performance time was not significantly influenced by a different, more demanding border design. 

Based on our preliminary study using paper prototyping, we expected participants to efficiently adopt the adaptive border design. 
Yet, we were surprised by how well participants adopted the technique, which is also reflected by the measured respective task performance time. 
In summary, there is no clear disadvantage of the adaptive border design, but the advantage of increased awareness of the data space dimensions. 
The additional step of first counting back the off-screen space and only then being able to decide on the position of off-screen points seems effortless. 

The adaptive border design preserves the surrounding data space similar to Focus-plus-Context (F+C) systems. 
One well-known F+C technique is the distortion lens: It allows the user to pick a specific part of the overall information landscape and then to magnify this part without losing context of the surrounding. 
Besides its merits, it maximizes the focus region in image space making judgments based on the data difficult, because the data representation is distorted together with the surrounding. 
We argue to use off-screen visualization with an adaptive border design if the data is paramount. \\

\noindent \emph{\textbf{DC2:} Use the adaptive border design for increased awareness of the data space. }

\begin{figure*}[t]
	\centering
	\includegraphics[width=1.0\linewidth]{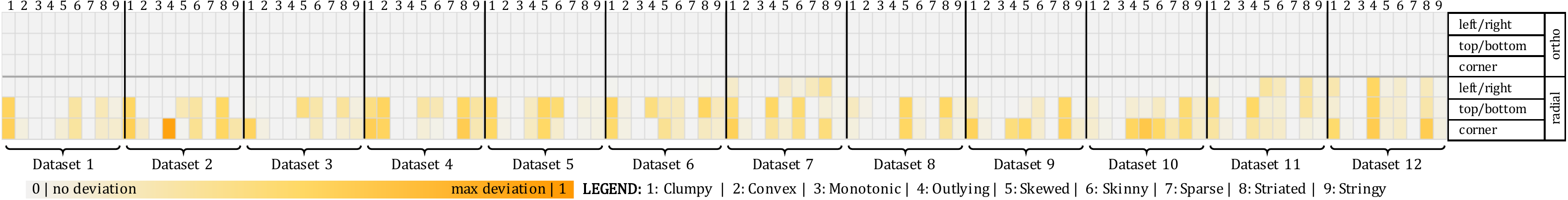}
	\caption{Systematic comparison between the orthographic (ortho) and radial projection strategy. We tested twelve datasets using Scagnostics regarding distortion of the sides left/right, top/bottom, as well as the corner region in a standard $16:9$ viewport. The table shows the deviation to the undistorted dataset and suggests to use the orthographic projection.}
	\label{fig:eval_scagnostics}
\end{figure*}

\subsection{Task Dependency}
Our experiment was not based on a certain task. 
After the study, 5 out of 18 participants reported on having difficulties at the beginning of the study to decide on a projection strategy. 
This is due to the fact that we solely presented projections integrated into the border, which could not be associated with a specific projection strategy.  
However, the encountered difficulties raise the question for task-dependency when considering the projection strategy: 
Can a projection strategy be associated with a certain task and thus eliminate ambiguities? 
Consider a navigational task such as using the TomTom~\footnote{https://www.tomtom.com/} device in your car. 
Typically, your position marks the point of interest on the map and a visual cue (normally an arrow) indicates the direction to drive to. 
A radial projection is used originating from your position and we all are used to it. 
But then, consider a monitoring task such as monitoring hospitals and their capacities in off-screen space due to a natural disaster. 
If not indicated otherwise, the hospitals share the same importance level and thus do not directly relate to a point of interest within the viewport. 
In this case, and because participants of our study intuitively used this strategy, the orthographic projection makes sense to apply. 
In summary, our results show that users did intuitively choose the orthographic projection strategy, however, the task at hand can influence our design for the consumer. \\

\noindent \emph{Based on the task at hand,}\\
\emph{\textbf{DC3:} apply radial projection for navigational tasks. }\\
\emph{\textbf{DC4:} apply orthographic projection for monitoring tasks.} \\

\looseness=-1
\noindent We are aware of the lack of evidence for \textbf{DC3} and \textbf{DC4}. However, particularly in systems where the direction plays a major role like navigation systems or computer games, the radial projection is applied. Despite discussions in related work, such use cases heavily relate to graph representations and along the edge routing. We consider it as natural to apply the radial projection for a point-to-point navigational task. Yet, the results show a clear preference for the orthographic strategy when no reference point within the viewport is given.

\subsection{Preservation of Data Characteristics}
Our results show that users favor the orthographic projection strategy for topology-preserving off-screen visualizations. 
However, compressing off-screen information into the border region is based on the distortion of distances. 
This raises the question whether the orthographic projection also preserves spatial data characteristics as accurate as possible. 
We, therefore, conducted a systematic evaluation using Scagnostics (scatterplot diagnostics) by Wilkinson et al.~\cite{DBLP:conf/infovis/WilkinsonAG05} to compare the radial and the orthographic projection.
Scagnostics are measures which indicate data-specific characteristics such as outliers (outlying), density (clumpy, skewed, sparse, striated), form (convex, skinny, stringy), and connection of points (monotonic). This approach connects all data points by a minimum spanning tree and applies predefined metrics to it.
We conducted a computational study using Scagnostics to get an idea of which projection strategy preserves the relations in the data best. We used similar ground-truth data sets to \cite{DBLP:conf/infovis/WilkinsonAG05} and moved the data off-screen, so that the data was projected to the dedicated border region.
For each dataset and Scagnostics measure, we computed the deviation to the undistorted dataset measures. 
For each projection strategy, we further considered the distortion for the sides left/right, top/bottom, as well as the corner region in a standard $16:9$ viewport. 
Since all values are normalized using feature scaling, the size of the border does not interfere with the results. 
Figure~\ref{fig:eval_scagnostics} shows the results. The table reveals that there is no deviation for the orthographic projection. 
One valid way to interpret this finding is that points are distorted uniformly in $x$ and $y$ direction which is why the Scagnostics measures remain unchanged due to feature scaling. 
The results of this systematic evaluation back up our results of the user study. In conclusion, the orthographic projection strategy is not only favored by users, but also preserves the data characteristics. 

\begin{figure}[h]
	\centering
	\includegraphics[width=0.99\linewidth]{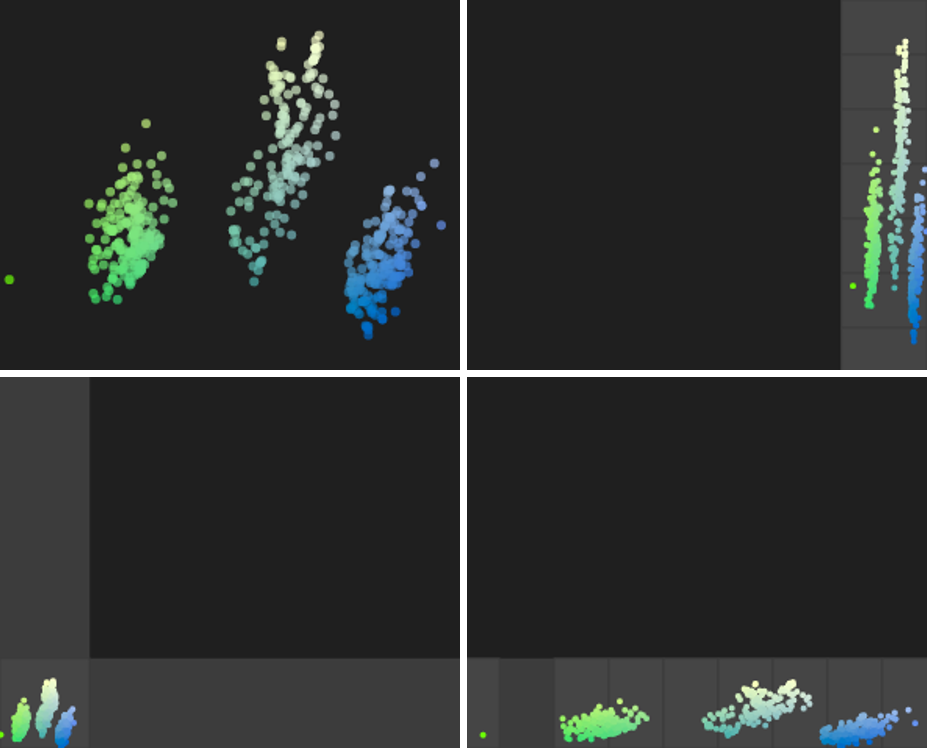}
	\caption{Example of a scatterplot using the adaptive border design and the orthographic projection strategy for approximately 500 data points. Top left: Overview of the dataset. Top right: The user panned the data into the off-screen space on the right hand side. The adaptive border was activated and presents all data points in the border on the right. Bottom right: The user panned the data into the lower off-screen space. Bottom left: In this case, all the data was panned to the off-screen corner area. Even though this area is the smallest compared to the vertical and horizontal extents, we perceive the general structure of the data.}
	\label{fig:example_scalability}
\end{figure}

\section{Limitations and Future Work}

The present paper has a number of limitations, which we aim to cope with in future work. 
Our evaluation aimed at pointing out which projection strategy meets the users' intuition in a topology-preserving off-screen environment as well as the effect of an adaptive border. \\

\noindent \textbf{Border Size:} We conducted our evaluation with a maximum border size similar to EdgeRadar~\cite{DBLP:conf/chi/GustafsonI07}. However, for the adapted border design the intrusion can decrease, raising the question: How big should the initial border size be? 
We assume that the dimensions of the border should be based on the amount of data being presented as well as the dimensions of the data space. 
Also, one can consider how much focus the user wishes for. 
In future work, we plan to investigate these factors with respect to the dimensions of the border region. 
Also, it can be of interest to apply the concept of an adaptive intrusion to other non-topology-preserving techniques such as Halo~\cite{DBLP:conf/chi/BaudischR03} or Wedge~\cite{DBLP:conf/chi/GustafsonBGI08}, this is, a non-uniform intrusion of visual cues dependent of the navigated space. \\

\noindent \textbf{Comparison to F+C:} Also, and as already mentioned, the awareness of the navigated space asks for an evaluation on data-basis against common Focus-plus-Context systems that use for example a lens-based approach. 
One idea is to apply both techniques to an abstract space, for example the two-dimensional result of a multivariate projection. 
Then, data points can be represented as multivariate data glyphs and users are asked to navigate that space based on a given data value. 
Both approaches actually show the data glyphs, one directly on the space, the other one projected along the display edge. 
A comparison between both reveals which technique to choose for fast data-based navigation. \\

\noindent \textbf{Applicability and Scalability:} 
The purpose of this study was to investigate which projection strategy meets the users' intuition in a topology-preserving environment, and whether the adaptive border impairs the perception. 
We conducted our study using a data space that is no more than 1.3x larger than the viewport including the off-screen visualization. The results of our study scale to data spaces larger than 1.3x the viewport: Imagine extending the data space to, for example, 10x larger than the viewport. An object positioned in the center of the off-screen space is still projected back to the center of the border extent, regardless of projection and off-screen space dimensions (illustrated in Figure~\ref{fig:distance_scale}). This is traced to the adaptive border that proportionally compresses the adjacent off-screen space and distances between objects. Task 1 tests the converse projection via a pointing hub forcing us to provide a restricted extent (as appeared in \cite{DBLP:conf/nordichi/MullerLHB14}). 

Another concern rises regarding the limited space in the corner regions when applying the orthographic projection strategy. At this point, we like to emphasize that an investigation of how well off-screen techniques scale to large datasets is not part of the contribution of this paper. Despite this, we like to show an example for a scatterplot including around 500 data points. Figure~\ref{fig:example_scalability} illustrates the result of an adaptive border intrusion as well as the orthographic projection applied to the scatterplot. While we agree that the amount of space reserved for the corner regions is limited, the overall structure of the data can still be perceived.

\section{Conclusion} 
The application of off-screen techniques to spatial information spaces has become prominent in recent years. However, one of the most important aspects has barely been considered so far, that is the preservation of the overall data topology as well as the effect of the projection strategy in such topology-preserving environments. In this paper, we proposed to add the data space awareness in addition to the topology preservation. 
We evaluated and showed that the application of an adaptive border does not hinder the overall performance accuracy as well as time. 
Furthermore, we showed that the orthographic projection strategy outperforms the radial projection strategy in off-screen visualizations, where the direction of the projected off-screen point is not implied.

\bibliographystyle{abbrv-doi}

\bibliography{sample}
\end{document}